\documentclass[reprint,aps, twocolumn, showpacs, preprintnumbers, superscriptaddress, amsmath, amssymb, prb]{revtex4-2}
\usepackage{amssymb}
\usepackage{mathrsfs}
\usepackage{amsmath}
\usepackage{graphicx}
\usepackage{epstopdf}
\usepackage{float}
\usepackage{multirow}
\usepackage{array}
\usepackage{ulem}
\usepackage{color}
\usepackage[colorlinks=true, letterpaper=true, pdfstartview=FitV, linkcolor=blue, citecolor=blue, urlcolor=blue]{hyperref}

\makeatletter
\newcommand{\Rmnum}[1]{\expandafter\@slowromancap\romannumeral #1@}
\makeatother
\usepackage{amsmath}

\begin{document}


\title{Efficient Quantum Mixed-State Tomography with Unsupervised Tensor Network Machine Learning\\}

\author{Wen-Jun Li}
\affiliation{
School of Physical Sciences, University of Chinese Academy of Sciences, Beijing 100049, China\\
}%



\author{Kai Xu}

\author{Heng Fan}
\affiliation{
Institute of Physics and Beijing National Laboratory for Condensed Matter Physics, Chinese Academy of Sciences, Beijing 100190, China\\
}%
\affiliation{
CAS Center for Excellence in Topological Quantum Computation, University of Chinese Academy of Sciences, Beijing 100190, China\\
}%

\author{Shi-Ju Ran}
\email{sjran@cnu.edu.cn}
\affiliation{Center for Quantum Physics and Intelligent Sciences, Department of Physics, Capital Normal University, Beijing 10048, China}

\author{Gang Su}
\email{gsu@ucas.ac.cn}
\affiliation{
 School of Physical Sciences, University of Chinese Academy of Sciences, Beijing 100049, China\\
}%
\affiliation{
 Kavli Institute for Theoretical Sciences, and CAS Center for Excellence in Topological Quantum Computation, University of Chinese Academy of Sciences, Beijing 100190, China
}%


\date{\today}

\begin{abstract}
Quantum state tomography (QST) is plagued by the ``curse of dimensionality'' due to the exponentially-scaled complexity in measurement and data post-processing. Efficient QST schemes for large-scale mixed states are currently missing. In this work, we propose an efficient and robust mixed-state tomography scheme based on the locally purified state ansatz. We demonstrate the efficiency and robustness of our scheme on various randomly initiated states with different purities. High tomography fidelity is achieved with much smaller numbers of positive-operator-valued measurement (POVM) bases than the conventional least-square (LS) method. On the superconducting quantum experimental circuit [Phys. Rev. Lett. 119, 180511 (2017)], our scheme accurately reconstructs the Greenberger-Horne-Zeilinger (GHZ) state and  exhibits robustness to experimental noises. Specifically, we achieve the fidelity $F \simeq 0.92$ for the 10-qubit GHZ state with just $N_m = 500$ POVM bases, which far outperforms the fidelity $F \simeq 0.85$ by the LS method using the full $N_m = 3^{10} = 59049$ bases. Our work reveals the prospects of applying tensor network state ansatz and the machine learning approaches for efficient QST of many-body states.
\end{abstract}

\maketitle


\section{\label{sec:1}INTRODUCTION}

In the noisy intermediate-scale quantum era~\cite{preskill2018quantum}, coherent manipulations of quantum states with tens or even hundreds of qubits can be realized on multiple platforms, including superconducting circuits~\cite{song201710,arute2019quantum,mooney2021whole}, trapped ions~\cite{smith2016many,friis2018observation}, and ultracold atoms~\cite{trotzky2012probing}. The fast-increasing scale of quantum computational platforms raises new challenges for quantum state tomography (QST)~\cite{vogel1989determination, leonhardt1995quantum, white1999nonmaximally, james2001measurement, Roos2004bell}. An essential obstacle is the so-called ``curse of dimensionality'' problem caused by the exponentially-increasing dimension of the Hilbert space~\cite{Paris2004}.

In the reconstruction of density matrices of quantum states from measurement data, both the dimension of Hilbert space and the complexity of QST increases exponentially with the system size. ~\cite{haffner2005scalable, lu2007experimental}. Meanwhile, exponentially-many computational resources to post-process the measurement data are required when using, e.g., the least squares (LS) method \cite{bardroff1996simulation,opatrny1997least} or the maximum likelihood estimation~\cite{leonhardt1995quantum,hradil1997quantum,banaszek1999maximum}.

Significant efforts have been made to overcome the ``curse of dimensionality''~\cite{granade2017practical,torlai2018neural,quek2021adaptive,ahmed2021quantum,cramer2010efficient,lanyon2017efficient,wang2020scalable}. For instance, neural networks, such as restricted Boltzmann machines, recurrent neural network, and adversarial neural networks, were employed for efficient QST~\cite{torlai2018neural,quek2021adaptive,ahmed2021quantum}. However, these models represent either classical probabilistic distributions or nonlinear mappings, which therefore fail to provide an intrinsic description of the quantum probabilities.

The close connections of tensor network (TN) to both ML and quantum many-body physics make it a uniquely suitable tool to study QST of many-body states. On the one hand, TN has achieved great successes in simulating quantum many-body systems~\cite{cirac2009renormalization,schuch2012resonating,bridgeman2017hand,orus2019tensor,ran2020tensor} and quantum information processing~\cite{verstraete2004density,verstraete2009quantum,dang2019optimising,pan2022simulation}. It serves as a compact representation of quantum many-body states, where the exponentially-many parameters can be safely reduced to be polynomial for the states obeying the area laws of entanglement entropy~\cite{eisert2010colloquium}. 

On the other hand, TN sheds light on developing novel quantum-inspired ML~\cite{stoudenmire2016advances, cohen2016expressive, levine2017deep, stoudenmire2018learning, han2018unsupervised, cheng2019tree, glasser2020probabilistic, sun2020generative, cheng2021supervised}. Quantum theories, such as entanglement and fidelity, were combined to perform ML tasks, including feature extraction, compressed sampling, and clustering~\cite{bengua2015optimal, cichocki2016tensor, cichocki2017tensor, ran2020tensorsa, martyn2020entanglement, liu2021entanglement, bai2022unsupervised, li2022non, PhysRevA.105.052424}. The recent works exhibited the prospects of TN, specifically matrix product state (MPS), for efficient QST~\cite{cramer2010efficient,lanyon2017efficient,wang2020scalable}, which, however, are limited to pure states. The efficient QST schemes for mixed states are urgently desired, particularly for the experimental investigations and applications. 

\begin{figure}
\includegraphics[width=1\linewidth]{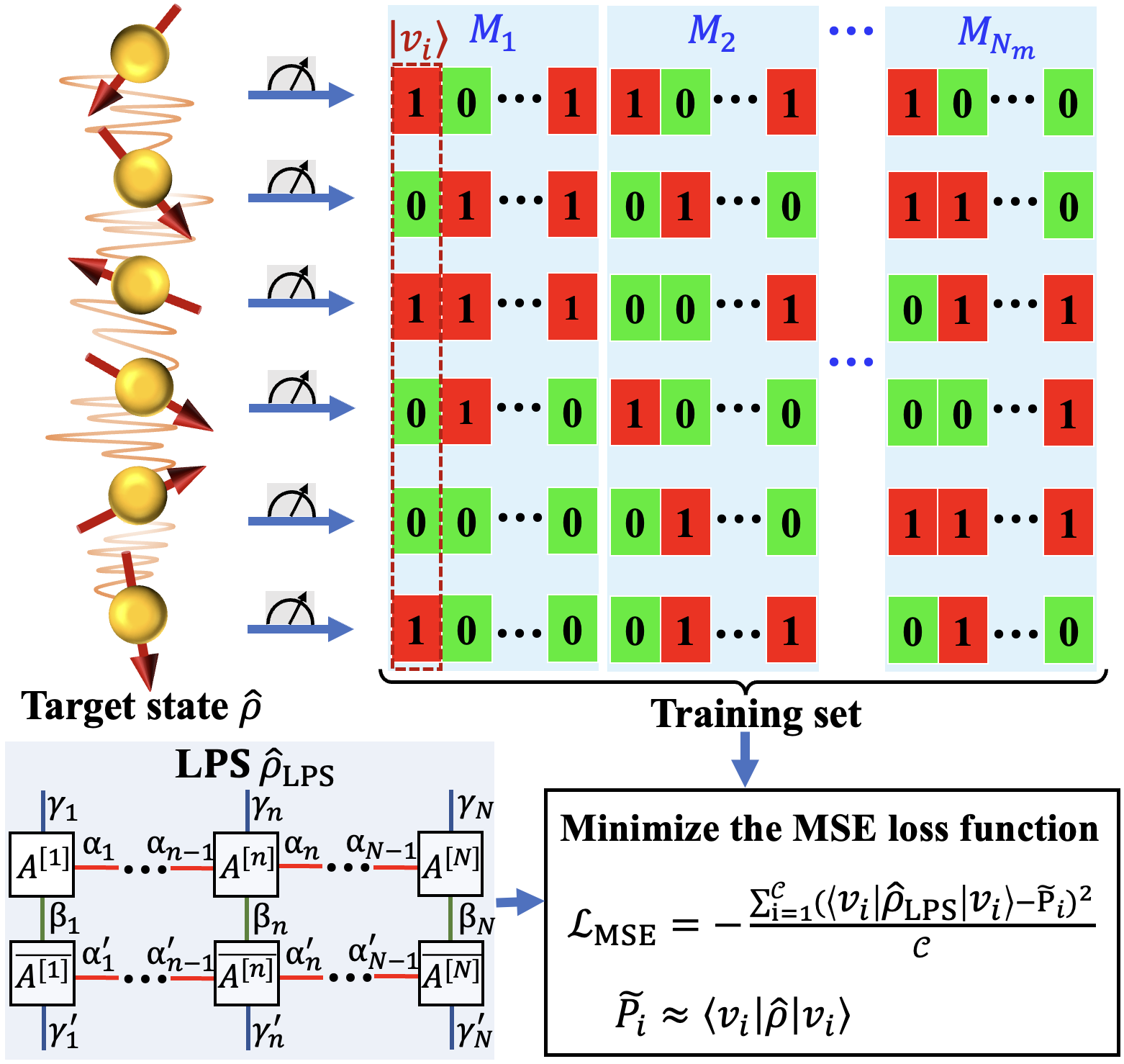}
\caption{\label{fig:flowchart} Diagram of the LPS-based QST scheme for an N-qubit mixed target state (denoted as $\hat{\rho}$). Initially, $\hat{\rho}$ is measured by $N_m$ randomly selected POVM bases to generate the training set with size $\mathcal{C}$. For each basis, there exist nearly $2^N$ different collapse states $| v_i \rangle$ with corresponding frequency $\tilde{P_i}$  obtained from sufficient measurement statistics. Subsequently, the parameters $A^{[n]}$ (n = 1, 2, $\ldots$, N) of the LPS model $\hat{\rho}_\text{LPS}$ are optimized by the gradient descent to minimize the MSE loss function $\mathcal{L}_\text{MSE}$, thereby complete the QST.}
\end{figure}

In this work, an efficient and robust mixed-state tomography scheme is proposed based on the locally purified state (LPS)~\cite{glasser2019expressive} (illustrated in Fig.~\ref{fig:flowchart}). LPS is a compact TN representation of mixed states whose complexity scales linearly with the system size. We here develop a LPS-based ML scheme to reconstruct the target state from the measurement data. For the QST of random states as the examples, we show a linear relation between the infidelity $F_\text{in}$ and $1/{\sqrt{N_m}}$ (with $N_m$ the number of randomly selected positive-operator-valued measurement (POVM) bases~\cite{nielsen2002quantum}), where the slope relies on the purity of the target state and the strength of noises. High fidelities are achieved by our scheme with small numbers of POVM bases.

We further consider the Greenberger-Horne-Zeilinger (GHZ) state~\cite{verstraete2005renormalization,fernandez2012gapless} that is experimentally prepared and measured on the superconducting circuit~\cite{song201710}. Our scheme achieves the fidelity $F \simeq 0.92$ with just $N_m = 500$, which surpasses the least-square (LS) method~\cite{bardroff1996simulation,opatrny1997least} using the complete $N_m = 3^{10} = 59049$ bases with $F \simeq 0.85$. Robustness to noises of our scheme is also demonstrated.

\section{\label{sec:2}Method}

LPS is a TN ansatz to model mixed states~\cite{glasser2019expressive}, where the total number of parameters scales linearly with the system size rather than exponentially. By choosing the eigenstates of Pauli-z operator $\prod_{\otimes n=1}^N {\hat{\sigma}}_z^n$ as the bases (i.e., $\{| s^n_z \rangle \langle s^n_z|\}$), an $N$-qubit LPS can be formulated as
\begin{widetext}
\begin{equation}
\hat{\rho}_\text{LPS} = \sum_{\alpha_1\alpha^{'}_1 \cdots \alpha_{N-1}\alpha^{'}_{N-1}}\sum_{\beta_1 \cdots \beta_{N}} A^{[1]}_{\gamma_1 \alpha_1 \beta_1}\overline{A^{[1]}_{\gamma^{'}_1 \alpha^{'}_1 \beta_1}} \cdots A^{[n]}_{\gamma_n \alpha_{n-1} \alpha_n \beta_n}\overline{A^{[n]}_{\gamma^{'}_n \alpha^{'}_{n-1} \alpha^{'}_n \beta_n}} \cdots A^{[N]}_{\gamma_N \alpha_{N-1} \beta_N} \overline{A^{[N]}_{\gamma^{'}_N \alpha^{'}_{N-1} \beta_N}} \prod^{N}_{\otimes n=1} | s^n_z \rangle \langle s^n_z|.
\label{eq-MPS}
\end{equation}
\end{widetext}
with $\overline{\boldsymbol{A}^{[n]}}$ the conjugate of $\boldsymbol{A}^{[n]}$. An illustration of LPS is given in Fig.~\ref{fig:flowchart}.

Here, $\{\gamma_n, \gamma^{'}_n\}$ (blue bonds in Fig.~\ref{fig:flowchart}) represent the physical indexes with the dimension $d = 2$ for qubits or spins-$1/2$'s; $\{\beta_n\}$ (green bonds) are the ``purification indexes'' with dim$(\beta_n)$= $d$ or $1$. The total dimension $D_{\beta} = \prod_{n} \dim(\beta_{n})$ (with $1 \leq D_{\beta} \leq d^N$) determines the upper bond of the degree of mixing of the LPS. By setting $\dim(\beta_{n})=1$ for all purification indexes (which is equivalent to removing all $\{\beta_n\}$ with $D_{\beta}=1$), the LPS is reduced to MPS and can only model the pure states. The indices $\{\alpha_n, \alpha^{'}_n\}$ (red bonds) are dubbed as virtual indexes, whose dimension $\chi$ determines the parameter complexity of the LPS~\cite{ran2020tensor}. The number of free parameters in an LPS scales linearly with system size $N$ as $O(Nd\mu\chi^{2})$, thereby avoiding the exponential complexity of the conventional QST schemes. It is expected that the LPS with a larger $\chi$ has stronger ability to learn information for, e.g., classification or generation~\cite{2023quantum,Charles2023}. 

To optimize the LPS for QST, we employ the standard POVM~\cite{nielsen2002quantum} consisting of three Pauli operators as measurement bases. The POVM bases of $N$ qubits are defined as
\begin{eqnarray}
\hat{M}  = \prod^{N}_{\bigotimes n=1} |s_{\alpha}^n\rangle \langle s_{\alpha}^n| \quad  \alpha\in(x, y, z),
\label{eq-measurement basis}
\end{eqnarray}
with $|s_{\alpha}^n\rangle$ the eigenstate of $\hat{\sigma}_{\alpha}^n$ of the $n$-th qubit. There are totally $3^N$ POVM bases.

The measurement with a chosen POVM basis will collapse the target state (denoted as $\hat{\rho}$) to a product state
\begin{equation}
| v \rangle = \prod^{N}_{\bigotimes n=1} | s_\alpha^n \rangle.
\label{eq-measurement results}
\end{equation}
There are in total $2^N$ possible results. The probability for obtaining a specific collapsed state by measuring on $\hat{\rho}$ satisfies $P(v)= \langle v  | \hat{\rho} | v \rangle$.

The training of LPS is inspired by the generative modeling of matrix product state~\cite{PhysRevX.8.031012} . We randomly choose $N_m$ POVM bases and implement measurement on sufficient copies of the target states. $N_c \simeq 2^N$ different collapsed states will be obtained for each POVM basis. There are in total $N_mN_c$ training samples in terms of ML. These samples are obtained by sampling numerically on the target state, or experimentally on the quantum platform. 

The loss function is defined as the mean-squared error (MSE)~\cite{james1992estimation}
\begin{equation}
	\mathcal{L}_\text{MSE} = - \frac{\sum^{N_m}_{m=1} \sum^{N_c}_{k=1}(\langle v^{[m,k]} | \hat{\rho}_\text{LPS} | v^{[m,k]} \rangle -  \tilde{P}^{[m,k]})^2}{N_m {N_c}},
	\label{eq-cost}
\end{equation}
with $|v^{[m,k]} \rangle$ the $k$-th collapsed state when measuring the target state $\hat{\rho}$ with the $m$-th POVM basis. The term $\langle v^{[m,k]} | \hat{\rho}_\text{LPS} | v^{[m,k]} \rangle$ is the probability of obtaining $| v^{[m,k]} \rangle$ by measuring the LPS model. This term can be calculated efficiently, whose cost scales linearly with the number of qubits $N$. 

The probability $\tilde{P}^{[m,k]}$ in Eq.~(\ref{eq-cost}) is estimated from the measurement data (i.e., training samples). With $\mathcal{L}_\text{MSE} \to 0$, the probability distribution predicted by the LPS approximately equals to that estimated from the measurement data. The tensors forming the LPS are updated by the gradient descent as $A^{[n]} \leftarrow A^{[n]} - \eta \frac{\partial \mathcal{L}_\text{MSE}}{\partial A^{[n]}}$ with $\eta$ the learning rate or gradient step. The gradients are obtained using the automatic differentiation technique of TN~\cite{bartholomew2000automatic,liao2019differentiable, torlai2020wave}.

\begin{figure}[b]
\includegraphics[width=1\linewidth]{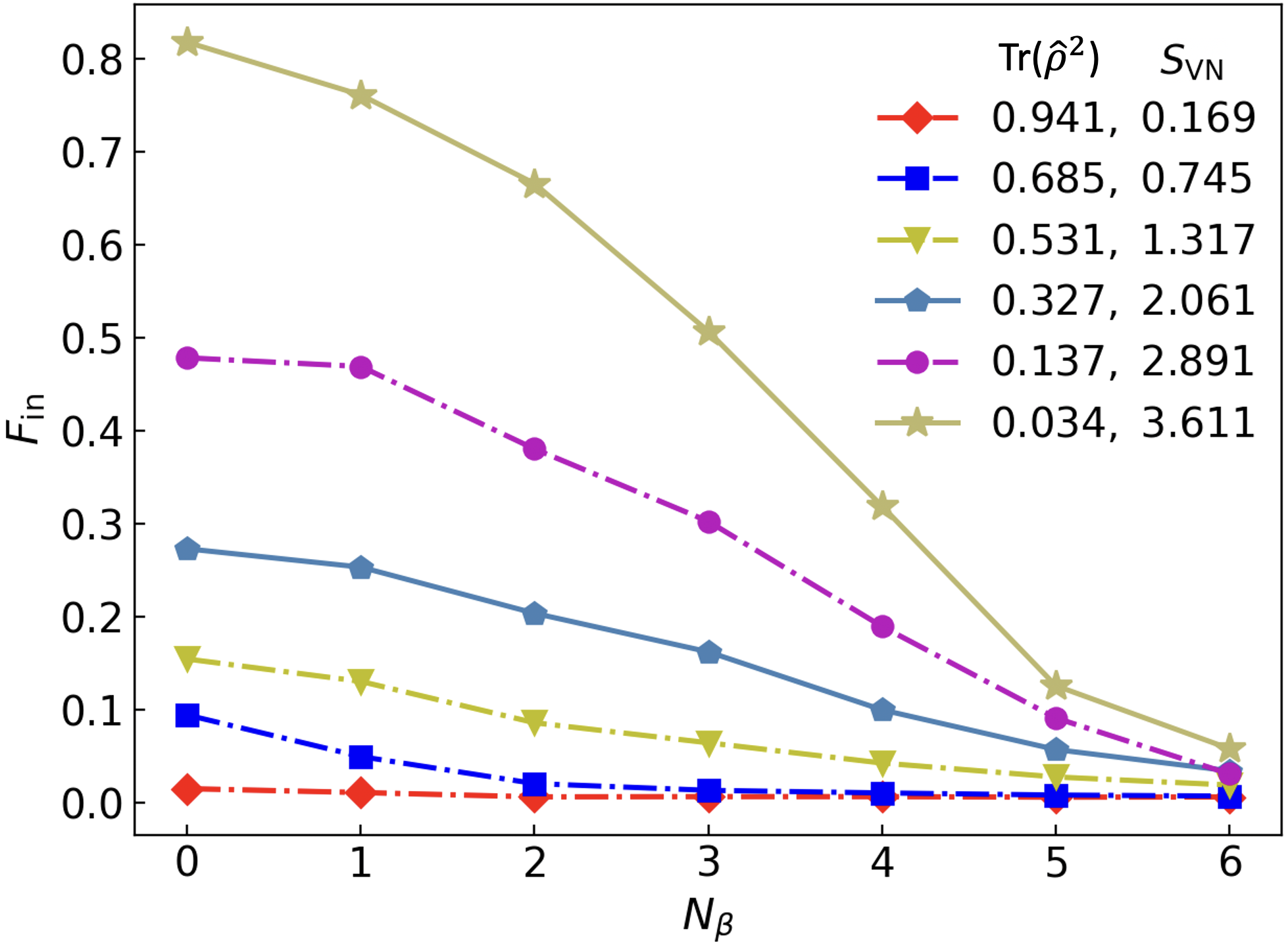}
\caption{\label{fig:Random mu} Scaling of the infidelity $F_\text{in}$ versus the number of $d$-dimensional purification index $N_\beta$ of the LPS model $\hat{\rho}_\text{LPS}$ for the target states $\hat{\rho}$ with varying degrees of mixing. The simulations are performed on the 6-qubit random states with different $\text{Tr}(\hat{\rho}^2)$~and~Von Neumann entropy $S_\text{VN}$. For each target state, we randomly select $N_m=300$ POVM bases and implement sufficient samples ($N_s=8192$) per basis. }
\end{figure}

\begin{figure}
\includegraphics[width=1\linewidth]{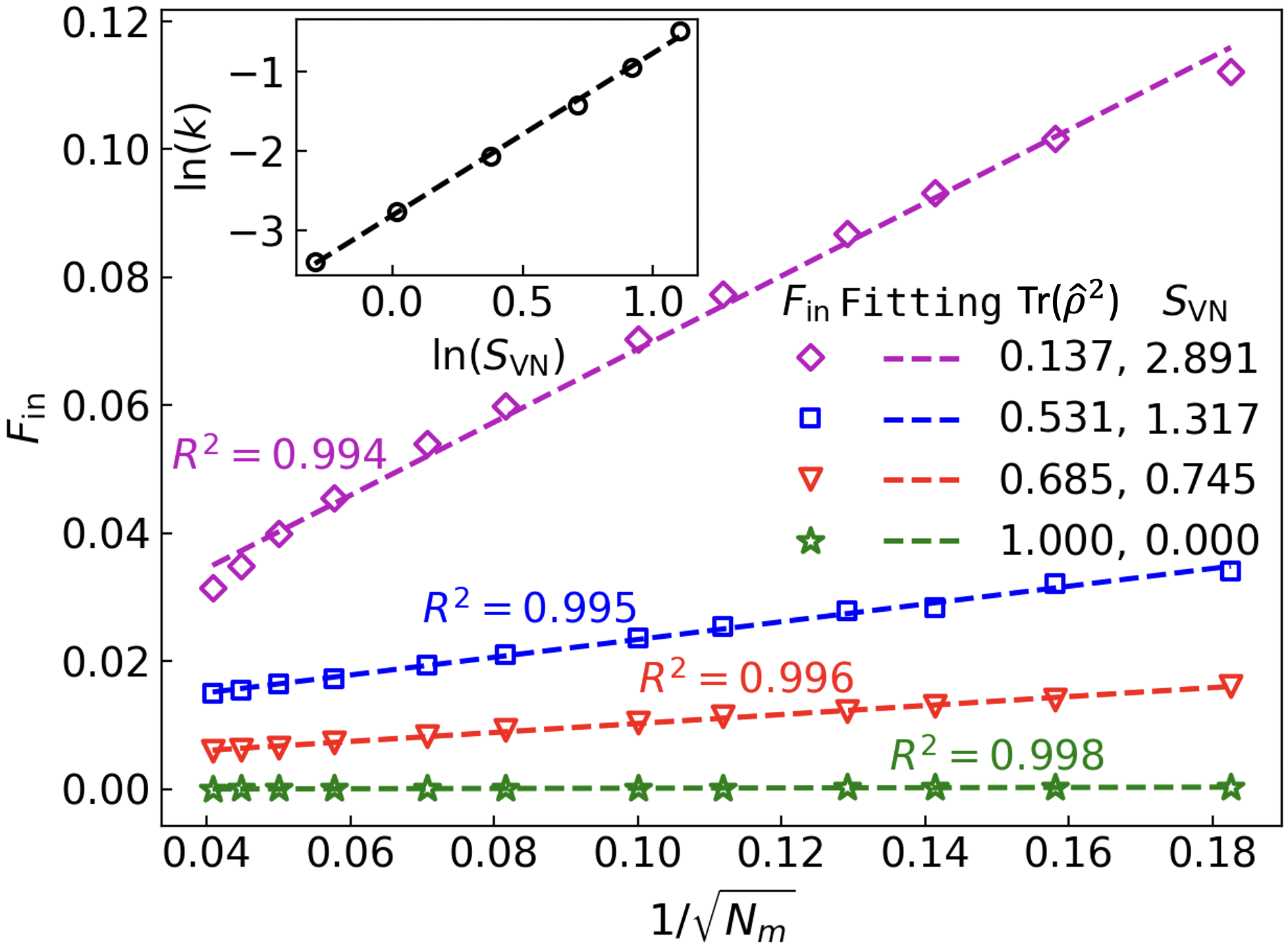}
\caption{\label{fig:Random} Efficiency of the LPS-based QST scheme for the target states with different purities (i.e. $\text{Tr}(\hat{\rho}^2)$~and~Von Neumann entropy $S_\text{VN}$). For each 6-qubit random state, we evaluate the efficiency of QST by the required number of POVM bases $N_m$. The results indicate a linear relationship between the infidelity $F_\text{in}$ and $1/{\sqrt{N_m}}$. For the target state with higher degree of mixing, the fitting difficulty increases, as evidenced by the decreasing $R^2$ coefficient and increasing slope. The inset shows the slope $k$ of the linear relation, which increases polynomially with $S_\text{VN}$ of the target state $\text{ln}(k)=2.03 \times \text{ln}(S_\text{VN})-2.81$, each point is the average of our scheme in 10 cases with different random seeds.}
\end{figure}

\begin{figure}[t]
\includegraphics[width=1\linewidth]{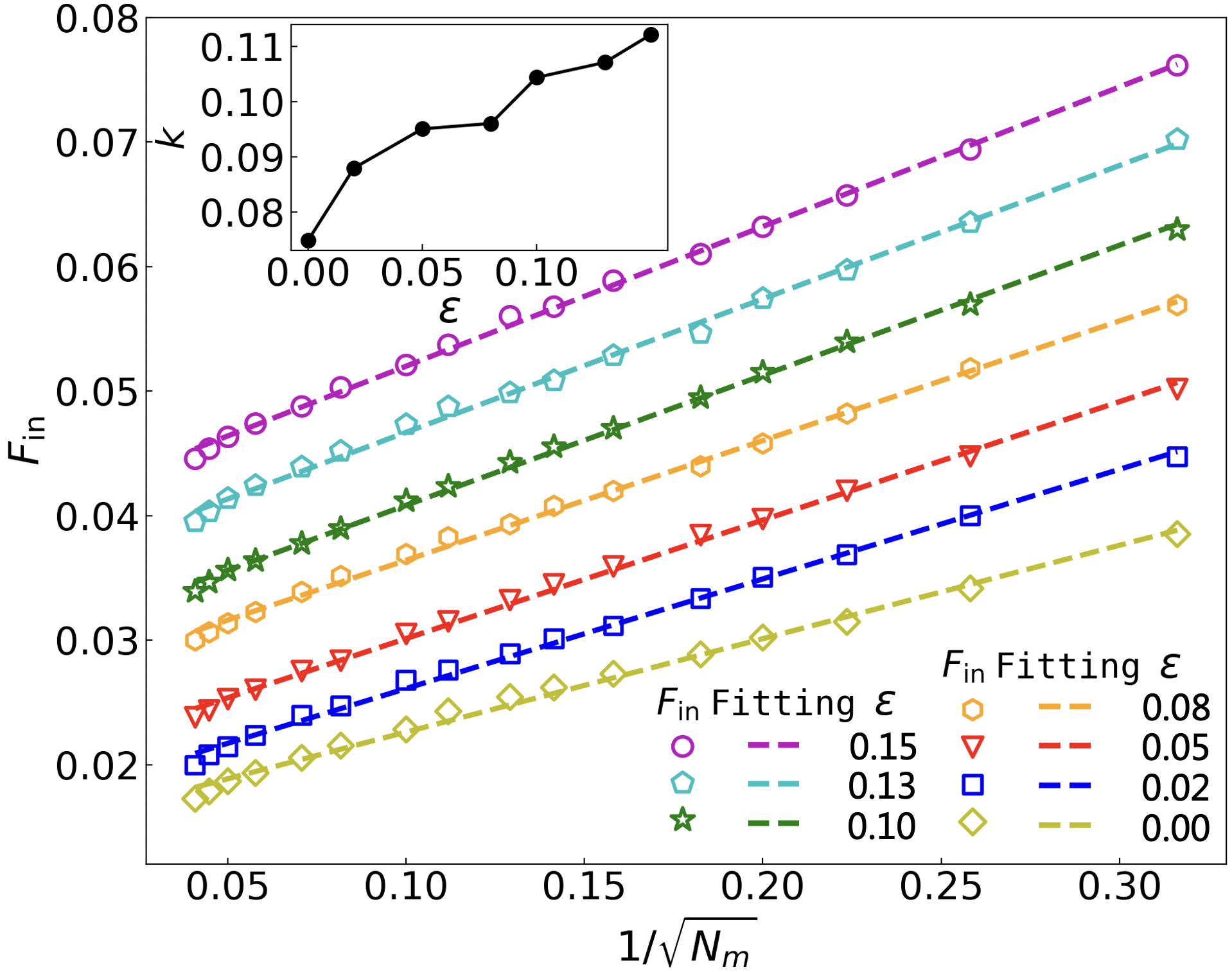}
\caption{\label{fig:Random noise} The robustness of the LPS-based QST scheme to the quantum depolarizing channel noise. The simulations are performed on the 6-qubit random states with $\text{Tr}(\hat{\rho}^2)=0.531$ and $S_\text{VN}=1.317$. The results show the infidelity $F_\text{in}$ still scales linearly with $1/{\sqrt{N_m}}$ even the measurement data suffer a certain intensity of noise. The inset shows how the slope $k$ varies with the noise level $\varepsilon$.}
\label{fig-randomMPS}
\end{figure}

\begin{figure}
\includegraphics[width=1\linewidth]{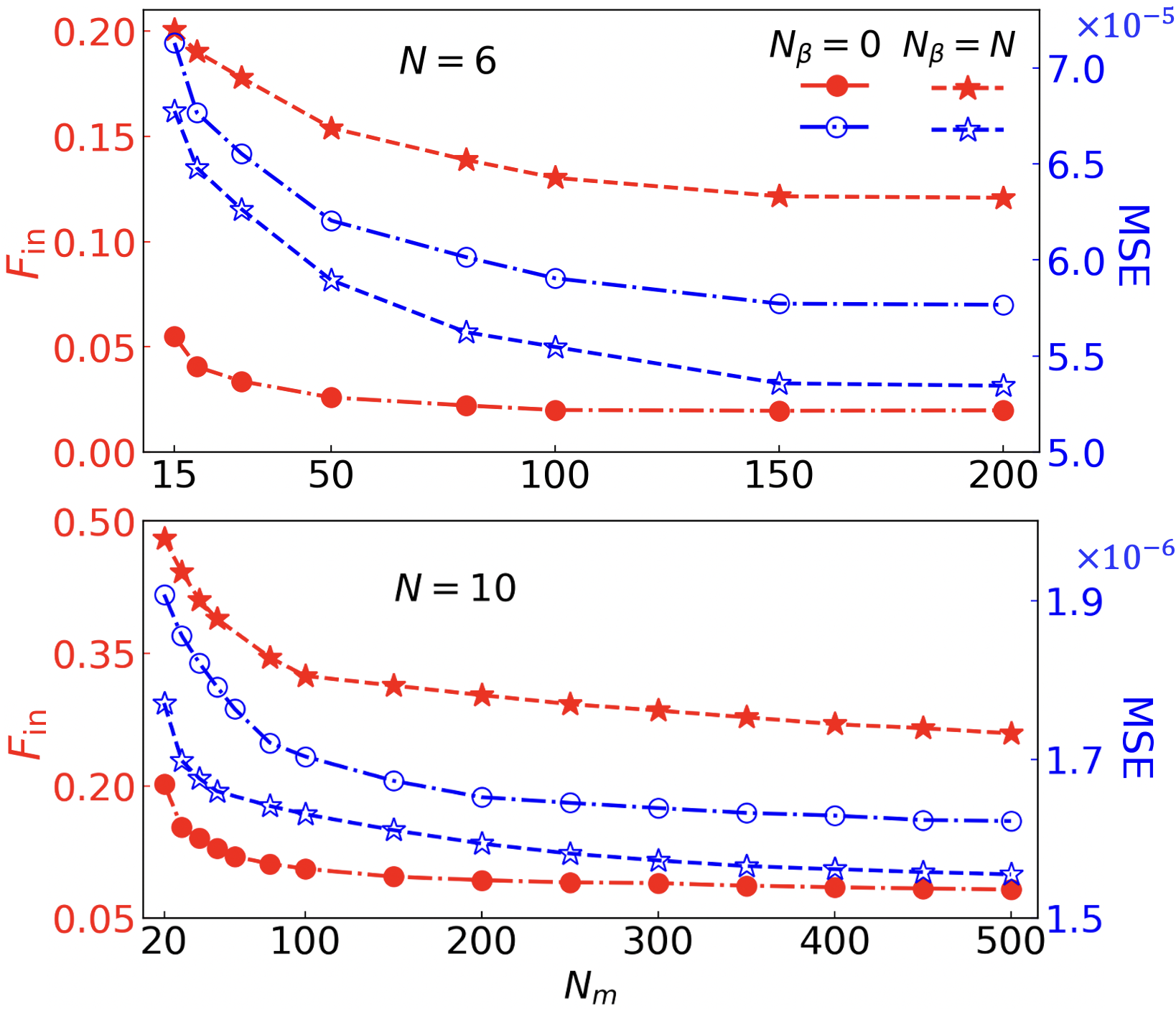}
\caption{\label{fig:GHZ noise} Scaling of the infidelity $F_\text{in}$ between the ideal GHZ state and the LPS, and MSE loss versus the number of POVM bases $N_m$ for the $N$-qubit $(N=6~\text{or}~10)$ GHZ states prepared and measured on the superconducting experimental circuit~\cite{song201710}. For an N-qubit GHZ state, our QST scheme is implemented using the LPS with $N_{\beta}=0$ and $N_{\beta}=N$, respectively.}
\end{figure}

\begin{figure*}
\includegraphics[width=1.0\linewidth]{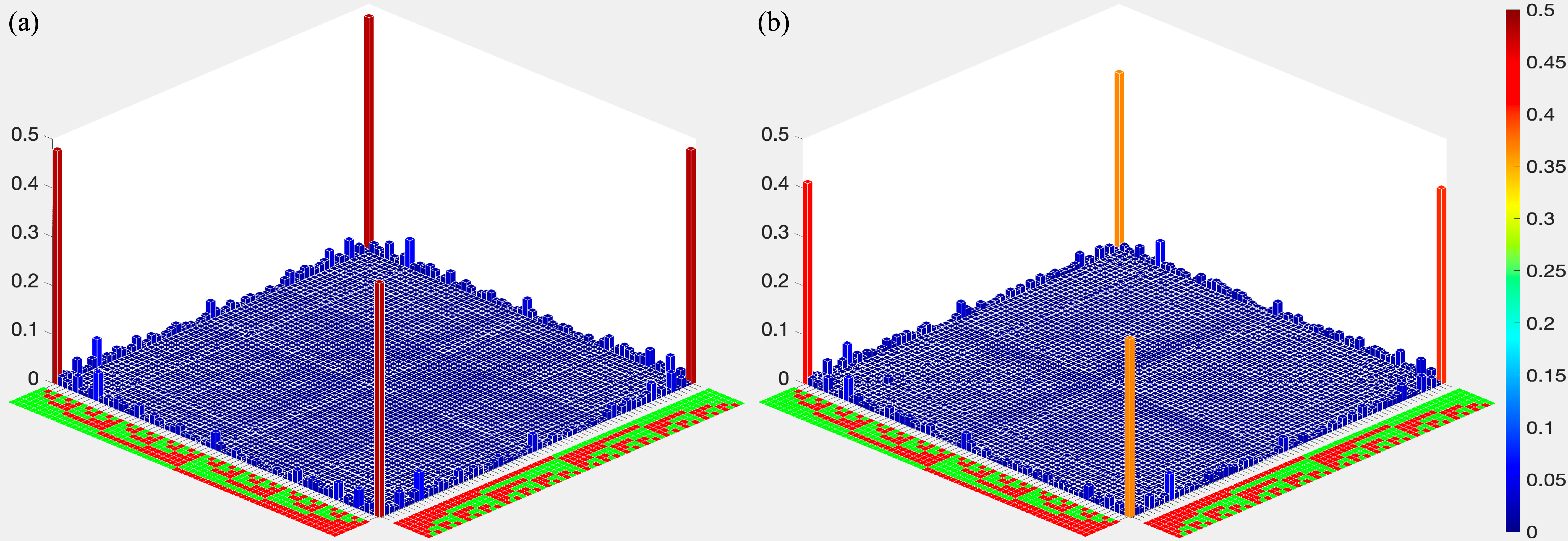}
\caption{\label{fig:GHZ} The probability distribution of the 6-qubit GHZ states obtained from our QST scheme using the LPS model with (a) $N_{\beta}=0$ and (b) $N_{\beta}=6$, respectively. The LPS models are optimized using a training set consisting of $N_m=200$ randomly selected POVM bases. The red and green squares represent the $|+\rangle$ and $|-\rangle$ bases, respectively.}
\end{figure*}

\section{\label{sec:3}Numerical benchmark}

The relation between the QST error and the purity of the target states is elusive due to the lack of efficient methods for many-body mixed states~\cite{PhysRevA.85.052107,qi2017adaptive,kazim2021adaptive}. Fig.~\ref{fig:Random mu} shows the infidelity $F_\text{in}$ versus the number of $d$-dimensional purification bonds $N_{\beta}$ for the QST of $6$-qubit random states with different purities. We take a small number of randomly select POVM bases ($N_m=300$ ), and for each basis we sample for sufficiently many times ($N_{s} = 8192$).

The infidelity $F_\text{in}$ is a widely-used measure of the distance between two quantum states with $F_\text{in} \in [0, 1]$. In our cases, it is defined as
\begin{equation}
	F_\text{in} = 1-\text{Tr}(\sqrt{\sqrt{\hat{\rho}}\hat{\rho}_\text{LPS}\sqrt{\hat{\rho}}}).
	\label{eq-Fidelity}
\end{equation}
One has $\hat{\rho} = \hat{\rho}_{\text{LPS}}$ for $F_\text{in}=0$.

The purity (or degree of mixing) can be characterized by $\text{Tr}(\hat{\rho}^{2})$ and von Neumann entropy $S_\text{VN}=-\text{Tr}(\hat{\rho}\ln \hat{\rho})$~\cite{nielsen2002quantum}.  For the pure states, one has $\text{Tr}(\hat{\rho}^{2})=1$ and $S_\text{VN}=0$. For a state close to be pure (say, with $\text{Tr}(\hat{\rho}^{2})=0.941$ and $S_\text{VN}=0.169$), low infidelities are achieved for any $N_{\beta}$. But as $\text{Tr}(\hat{\rho}^{2})$ decreases (or $S_\text{VN}$ increases), the infidelity increase fast when using the pure-state LPS ($N_{\beta}=0$, equivalent to MPS). 

Generally for the mixed states, the infidelity $F_\text{in}$ decreases with the purity, indicating that the QST will be more difficult as the target state becomes more mixed. Meanwhile, $F_\text{in}$ decreases with $N_{\beta}$, which is consistent with the fact that the ability of LPS on representing mixed states increases with $N_{\beta}$. These results demonstrate that significant error will appear when implementing the tomography of a mixed state by a pure-state model. The LPS with ``full'' purification indexes ($N_{\beta}=N$) exhibits the highest fidelity in spite of the purity of the target state. 

The efficiency of QST can be evaluated by the required number of POVM bases $N_m$. Fig.~\ref{fig:Random} shows the relation between $F_\text{in}$ and $1/\sqrt{N_m}$, which is linear for about $N_m \sim O(10^{1})$ or $O(10^{2})$. The $R^{2}$ coefficient, which characterizes the deviation from the linear fitting, just slightly decreases when the target state becomes more mixed. For a pure state, the slope of the linear relation is almost zero, showing that $O(10^{1})$ POVM bases are sufficient for an accurate QST using LPS. The inset demonstrates the slope $k$ of the linear relation, which increases polynomially with the von Neumann entropy $S_\text{VN}$ of the target state $\text{ln}(k)=2.03 \times \text{ln}(S_\text{VN})-2.81$. This again revealings the significant challenge on the QST for mixed states.  

\section{Tomography with noisy measurement data}

Realistic quantum systems are always subject to undesired interactions with the surrounding environment, which inevitably introduce noises. For instance, the measurement can be affected by the noises caused by imperfect perpetuation and operations on quantum states. To simulate with noisy data, we mix the target state with the depolarizing channel quantum noise~\cite{nielsen2002quantum} as,
\begin{equation}
\hat{\rho}_{\varepsilon} = (1 - \varepsilon)\hat{\rho} + \frac{\varepsilon}{2^N}I,
\label{eq-Fidelity}
\end{equation}
where $\varepsilon$ controls the strength of noise and $I$ is the identity. The training set is generated by sampling on $\hat{\rho}_{\varepsilon}$.

The robustness of our scheme is demonstrated by the infidelity $F_\text{in}$ between the LPS $\hat{\rho}_\text{LPS}$ and the ideal target state $\hat{\rho}$, where the LPS is trained by the measurement data from the noisy state $\hat{\rho}_{\varepsilon}$. As shown in Fig.~\ref{fig:Random noise}, the linear scaling of $F_\text{in}$ with $1/{\sqrt{N_m}}$ still persists in general under noise. Instead, the noisy just increases the intercept and slope (see the inset of Fig.~\ref{fig:Random noise}) of the linear relation.

To benchmark on a realistic quantum computer, we consider the prototypical $N$-qubit GHZ state prepared and measured on the superconducting experimental circuit~\cite{song201710}. The ideal GHZ state is pure and can be defined as
\begin{eqnarray}
| \phi_{\text{GHZ}} \rangle = (| +_1,...,+_N\rangle + e^{i\beta} | -_1,...,-_N\rangle)/\sqrt{2}, \label{subeq:2}
\end{eqnarray}
with $| \pm_n \rangle = (| 0_n \rangle \pm i^N e^{i\gamma_n}| 1_n \rangle)/\sqrt{2}$. The phase variation $\beta$ and the dynamical phase $\gamma_n$ ($n=1,...,N$) are determined experimentally.

We take the $6$-qubit and $10$-qubit experimentally prepared GHZ states $\hat{\rho}_\text{exp}$ as examples to demonstrate the efficiency and robustness of our QST scheme. We experimentally ensure that there are sufficient samples for each POVM basis to obtain stable frequency statistics. Our scheme is applied to reconstruct the $\hat{\rho}_\text{exp}$ using the LPS with $N_{\beta}=0$ (pure state with $D_{\beta}=1$) and $N_{\beta}=N$ (mixed state with $D_{\beta}=d^{N}$), respectively. 

Fig.~\ref{fig:GHZ noise} shows that the infidelity $F_\text{in}$ between the ideal GHZ state $| \phi_{\text{GHZ}} \rangle$ and the pure LPS is much lower than that between $| \phi_{\text{GHZ}} \rangle$ and the mixed LPS. The experimentally prepared state is mixed due to the noises. As we can design the number of purification indexes, the reconstruction is implemented by the pure-state space by taking $N_{\beta}=0$. This is analogous to ``projecting'' the noisy mixed state back to the pure state with minimal distance characterized by MSE [Eq.~(\ref{eq-cost})]. In this way, noises are suppressed by taking the pure LPS for QST. Specifically for the 10-qubit GHZ state, our scheme achieves an infidelity $F_\text{in} \simeq 0.08$ with just $N_m = 500$ bases, which far outperforms the LS method using the full $N_m = 3^{10} = 59049$ bases with $F_\text{in} \simeq 0.15$.

If we faithfully attempt to reconstruct the noisy state, which is mixed, the mixed LPS will surely show higher accuracy. This is demonstrated in Fig.~\ref{fig:GHZ noise}, where the MSE of the mixed LPS is lower than that of the pure LPS. Note the MSE is evaluated from the measurement data on the noisy experimental GHZ state $\hat{\rho}_\text{exp}$.

Fig.~\ref{fig:GHZ} demonstrates the probability distribution of the $\hat{\rho}_\text{LPS}$ with $N_\beta=0$ and $N_\beta=6$, respectively, for the QST of the $6$-qubit experimental GHZ state (taking $N_m=200$). The bases are illustrated as the red ($|+\rangle$) and green ($|-\rangle$) squares. The largest probabilities appear at the four corners of the ``panel''. The fluctuation of the outer ring aligns with the imperfect experimental GHZ state. This is due to the imperfections of the Hamiltonian for generating the GHZ state, including the inconsistent coupling strengths between qubits and the approximate dispersion coupling between the qubit and the cavity. The pure LPS suppresses the fluctuations, particularly those within the panel, thus resulting in a lower infidelity between the LPS and the ideal GHZ state. In comparison, larger fluctuations  appear in the probabilistic distribution of the mixed LPS.

\section{CONCLUSION}

In this work, we propose an efficient and robust quantum state tomography (QST) scheme for mixed states by combining the local purified state (LPS) with unsupervised tensor-network machine learning. The LPS lowers the exponential complexity of QST to be just linear to the number of qubits. The linear relation between the infidelity and the number of measurement bases is exhibited, where we reveal a polynomial dependence of the slope against the von Neumann entropy of the target mixed state. Robustness to noises is demonstrated with the measurement data from numerical simulations and from experiments on the quantum circuit~\cite{song201710}. Our work uncovers the wide prospects on applying powerful TN ansatz and the quantum-inspired machine learning algorithms for efficient tomography of quantum many-body states.

\begin{acknowledgments}
We acknowledge discussion with Zheng-zhi Sun at early stages of this project. This work is supported in part by the NSFC (Grant No. 11834014 and No. 12004266), Beijing Natural Science Foundation (Grant No. 1232025), the National Key R\&D Program of China (Grant No. 2018FYA0305804), the Strategetic Priority Research Program of the Chinese Academy of Sciences (Grant No. XDB28000000), and Innovation Program for Quantum Science and Technology (No. 2021ZD0301800).
\end{acknowledgments}

\appendix

\nocite{*}

\bibliography{Ref.bib}

\end{document}